 \documentclass[preprint2]{aastex}

\slugcomment{Millennium Essay} 

\shorttitle{Pagel} 
\shortauthors{Chemical evolution of galaxies}

\begin{document}


\title{Chemical Evolution of Galaxies}  

\author{B.E.J. Pagel} 
\affil{Astronomy Centre, CPES, University of Sussex,  
    Brighton BN 1 9QJ, UK}

\begin{abstract}
Chemical evolution of galaxies brings together ideas on stellar evolution 
and nucleosynthesis with theories of galaxy formation, star formation and 
galaxy evolution, with all their associated uncertainties. In a new 
perspective brought about by the Hubble Deep field and follow-up investigations 
of global star formation rates, diffuse background etc., it has become 
necessary to consider the chemical composition of dark baryonic matter as 
well as that of visible matter in galaxies.  
\end{abstract}




\section{Introduction}

The seeds of an idea of galactic chemical evolution were planted by Sir 
Fred Hoyle a long time ago (Hoyle 1946). In the ensuing half century 
those seeds have grown and proliferated like a cashew-nut tree, with many 
roots and branches, some firmer than others; but it is not a mature 
subject in the sense that, say, stellar evolution is, 
with the basic ideas well understood and steady progress being made on 
the basis of previous knowledge.  There is still a lot of guesswork 
involved in 
the physics of star formation, and the 
very origin of galaxies like our own depends on an as yet unknown 
balance between monolithic collapse (Eggen, Lynden-Bell \& Sandage 1962), 
accretion of dwarf galaxies (Searle \& Zinn 1978), hierarchical clustering 
(White \& Rees 1978), mergers (Toomre 1977), inflows and 
outflows. These issues were already raised in the classic conference 
proceedings edited by Tinsley \& Larson (1977).      

\section{Ingredients of chemical evolution models}

A chemical evolution model needs to put together at least 5 ingredients: 

\subsection{Stellar yields} 

Starting with 
Arnett (1978) and Renzini \& Voli (1981), there 
have been numerous systematic investigations of stellar element production 
and ejection, as a function of the initial mass and chemical composition of 
the star.  The broad outlines are clear, but not the details: 
massive stars explode as core-collapse 
supernovae,  but above some mass limit, which could be of 
the order of $50M_{\odot}$, the outer layers may fall back into a black hole, 
reducing or 
eliminating ejection into the interstellar medium (ISM). Such stars 
will, however, have ejected significant amounts of helium and carbon at 
earlier stages in stellar winds (Maeder 1992).  Uncertainties arise from 
the $^{12}$C ($\alpha,\gamma)^{16}$O  
reaction rate, the treatment of convection and mass loss, the explosion mechanism 
and the mass cut, which is put in by hand to get a $Y_{e}$-value appropriate for the 
observed composition of the iron group 
(Woosley \& Weaver 1995; Thielemann et al.\ 1996).  
The bulk of the iron group (say 2/3 in the 
Solar System) comes, however, from thermonuclear supernovae, Type Ia, consisting of 
a white dwarf that explodes after accreting matter from a companion 
(Thielemann et al.\ 1986). Currently, departures from spherical symmetry are being 
investigated. 

Similar uncertainties apply to intermediate-mass stars, which are responsible for 
a significant part of nitrogen and $^{13}$C and for the main s-process
(van den Hoek \& Groenewegen 1997; Marigo, Bressan \& Chiosi 1998; Gallino et al.\ 
1998).             

Talbot \& Arnett (1974) introduced the distinction between `primary' and `secondary' 
nucleosynthesis products, according as the yields were insensitive, or sensitive, to 
the composition of the progenitor star. 
There is little abundance evidence for `secondary' 
behaviour among many elements for which it was once expected (e.g. s-process), but 
carbon displays secondary-like behaviour, not because of its nuclear progenitors but 
because higher metallicity favours stronger stellar winds (Gustafsson et al.\ 1999). 
Nitrogen, while behaving as a primary element in low-metallicity H~{\sc ii} regions, 
shows a gradually increasing N/O ratio that finally increases even more steeply 
than a secondary element, because of its dependence on the quasi-secondary carbon 
(Henry et al. 2000). 

\vspace{-1mm} 
\subsection{The initial mass function} 

The overall yield from a generation of stars\footnote{Defined as the mass of 
elements freshly produced and ejected by a generation of stars, divided by 
the mass remaining as long-lived stars or compact remnants (Searle 
\& Sargent 1972).}  depends on the initial mass function 
(IMF), first investigated by Salpeter (1955).  Many references to the Salpeter 
function 
nowadays 
refer explicitly or implicitly to a function with the Salpeter slope extending to  
$0.1M_{\odot}$  at the low-mass end, which is neither accurate nor any part of what 
Salpeter originally claimed.  While investigations of field stars in the solar 
neighbourhood have led to significantly steeper functions at the high mass end 
(
Scalo 1986), extragalactic studies almost invariably 
confirm Salpeter's slope above $1M_{\odot}$ or so (e.g.\ Madau et al.\ 1996). This 
leads to some intriguing 
consequences for galactic chemical evolution models, as the full Salpeter 
function (extending between 0.1 and 100$M_{\odot}$, say) leads to an overall yield 
around 2$Z_{\odot}$, too high for the solar neighbourhood; modellers using that 
function then either adopt a still lower low-mass truncation and/or assume
an upper limit of 50$M_{\odot}$ or less to stars that become supernovae, more 
massive stars  
locking themselves in black holes. The Miller-Scalo and Scalo functions do not need 
this device, but their lower overall yield has a problem explaining the metallicity 
of X-ray gas in clusters of galaxies. 

Is the IMF invariable? As there is no real theory, the question is wide open, 
but it is of interest to explore how much can be explained on the basis that 
it is, apart from random realizations of an underlying universal function. 
In this spirit, Pagel \& Tautvai\v sien\. e (1998) have attempted to model 
the chemical evolution of the Magellanic Clouds on the basis 
of identical yields to those prevailing in the solar neighbourhood  
(regardless of what 
particular combination of stellar yields and IMF is responsible for them), 
rather than blame their low metallicities on a steeper or more bottom-heavy IMF. 
Observations tend to favour a universal Salpeter slope above some critical mass 
below which it flattens or turns over; 
that critical mass may or may not be variable (Elmegreen 2000). 

\vspace{-1mm} 
\subsection{Star formation rates} 

Schmidt (1959) proposed a star formation law depending on a power between 1 
and 2 of the volume or surface density of gas; such laws have been used in many 
models and can give a 
good account of the distribution of gas density and abundances in the Milky Way 
(e.g.\ Matteucci \& Fran\c cois 1989), especially when some form of self-regulation 
is incorporated in the coefficients (Dopita \& Ryder 1994) . In dwarf and starburst 
galaxies, on the other hand, 
star formation often occurs in sporadic bursts, perhaps involving both negative 
and positive feedback mechanisms.  Kennicutt (1998) has given observational 
evidence for an overall correlation 
of star formation rates with the surface density of H~{\sc i}, with a definite 
threshold of order a few $M_{\odot}$ pc$^{-2}$ which may be related to dynamical 
stability criteria, and this idea has been used by Chiappini, Matteucci \& 
Gratton (1997) to account for the hiatus in star formation that appears to have 
occurred between the formation of the thick and thin disks.   

\vspace{-1mm}  
\subsection{Stellar populations}

One issue that has to be addressed, most notably in modelling the Milky Way, is 
the relationship between different stellar populations --- the halo, the bulge, 
the thick disk and the thin disk. To what extent have they evolved concurrently, 
either in space or in time, successively or independently?  Partly because of 
angular momentum considerations (Wyse \& Gilmore 1992), 
opinion has veered away from the older idea of a temporal succession: halo, 
thick disk, thin disk (Burkert, Truran \& Hensler 1992) towards the view that 
the halo and disks evolved independently, gas lost from the halo ending up in 
the bulge or the intergalactic medium. The thick disk is old and preceded the 
thin one, but with a considerable hiatus (Fuhrmann 1998), either because of 
the above-mentioned threshold effect or because of a merger which led to 
the thickening of the disk in the first place. 

\vspace{-1mm} 
\subsection{Interaction with other galaxies and the intergalactic medium}

Since the pioneering paper by Larson (1972), it has become clear that inflow 
of relatively unprocessed material is potentially an important factor, notably 
in helping to solve the notorious G-dwarf problem (see below), and it was 
also Larson who developed models of terminal galactic winds to account for the 
luminosity-metallicity relation and predicted 
the presence of heavy elements in intra-cluster gas (Larson \& Dinerstein 1975).  
More recent `chemo-dynamical' models also take into account the 
multi-phase structure of the ISM, with stellar ejecta supplying the 
hot medium and fresh stars forming in the cool one (Samland, Hensler \& Theis 
1997). 




\vspace{-1mm} 
\section{What have we learned from observations?}

\subsection{The G-dwarf problem} 

Sometimes dismissed as a little local difficulty, the G-dwarf problem 
(van den Bergh 1962; Schmidt 1963; Pagel \& Patchett 1975; Lynden-Bell 
1975) has proved 
to be a severe constraint on chemical evolution models, not only in 
the solar neighbourhood, but in elliptical galaxies (Bressan et al.\ 1994; 
Worthey et al.\ 1996) and the Magellanic Clouds (Cole et al.\ 2000) as well. 
The problem is that in all these cases there is a narrow distribution 
of metallicity (MDF), whereas naive concepts of chemical evolution lead to 
the expectation of a broad one. Such a narrow distribution probably 
helped to hold up the abandonment of the idea of a universal cosmic 
abundance distribution (cf.\ Sandage 2000), and it also explains why 
stellar population synthesis models assuming just a single metallicity 
(SSPs) have been quite successful --- more so than models incorporating 
chemical evolution up to now.  These models are gradually becoming more 
refined, often with an indication of  
a bimodal metallicity distribution (Maraston \& Thomas 
2000), which may be understandable as a consequence of  
mergers. Closely related to the G-dwarf problem  
is the lack of a single clear age-metallicity relation in the solar 
neighbourhood, explainable only in part by the mixing of populations 
from different galactocentric distances evolving on different 
time-scales (Edvardsson et al.\ 1993). 

The MDF is broader in the Galactic bulge and broader still in the 
halo, with an apparently higher yield (at least for $\alpha$-elements) 
in the former case and a lower one in the latter, where a modified 
Simple model assuming outflow actually fits the MDF rather well 
(Hartwick 1976). If the outflow went into the bulge, that might 
give an explanation for its higher apparent yield on the lines of 
the `concentration model' of Lynden-Bell (1975) and the models of 
elliptical galaxy formation by Larson (1976). The halo MDF is becoming 
well known from the heroic efforts of Beers et al.\ (1998), and it fits 
the modified Simple model down to about [Fe/H] $\simeq -3$; below 
that it falls short and below $-4$ there are 2 stars or less when 
nearly 10 might have been expected. If significant, this discrepancy 
could indicate the presence of a distinct Population III of massive 
stars only, or it could merely be the result of low-mass stars being 
formed in the neighbourhood of exploding supernovae, for which there is 
other evidence (see below).          

\vspace{-1mm} 
\subsection{Abundance patterns}

Abundance ratios are a better `clock' than metallicities 
themselves (however defined). The `$\alpha$-rich' effect (Wallerstein 
1962) and the O/Fe enhancement (Gasson \& Pagel 1966; Conti et al.\ 1967)
are a steady function of metallicity in the thin disk, reaching more 
or less a plateau in the thick disk and halo, and attributed to the 
diminishing contribution of SNIa to the elements in increasingly 
old stars (Wheeler, Sneden \& Truran 1989). There is currently 
controversy as to whether O/Fe actually has a plateau or rises 
steadily with diminishing Fe/H (Israelian et al.\ 1999; Boesgaard et 
al.\ 1999; Fulbright \& Kraft 1999). Numerical GCE models predict a 
steeper rise in ratios like [Mg/Fe] than is observed, but this 
depends on assumptions about SNII yields that may be invalid 
and the predicted slope is reduced in any case when finite mixing 
times are taken into account (Thomas, Greggio \& Bender 1999). 
Complications in this pattern have been found in two respects: (i) 
some halo stars have more solar-like $\alpha$/Fe ratios than do 
thick-disk stars and other halo stars at the same Fe/H (Nissen \& 
Schuster 1997), maybe because they came from more slowly evolving 
dwarf galaxies like the Magellanic Clouds, which show a similar 
pattern; and (ii) the $\alpha$-rich pattern persists among thick-disk 
stars right up to solar metallicity, indicating a fast-evolving 
`get rich quick' population, which may extend into the bulge,\footnote 
{The `get-rich-quick' nature of the M 31 bulge was already noted by 
Baade (1963), as Rich (2000) has recently reminded us.} 
and a hiatus with no star formation, just delayed iron-group production 
combined with some dilution of overall metallicity, before the first 
stars of the thin disk were formed (Fuhrmann 1998; Gratton et al.\ 
2000).          

The time-delay model for $\alpha$/Fe effects comes up against some 
difficulties in the case of elliptical galaxies, where there is a 
very well-marked correlation between Mg$_ 2$ and velocity dispersion, 
but a less well marked one for iron features (Worthey, Faber \& 
Gonzales 1992).  This should imply a faster star formation time-scale 
for larger galaxies, which is hard to understand on the basis of 
either monolithic or hierarchical clustering models (Thomas \& 
Kauffmann 1999). 
 
At very low metallicities like [Fe/H] $\simeq -3$, just where the 
modified Simple model MDF is breaking down, new abundance patterns 
emerge, with a large scatter in r/Fe and other ratios, indicating the 
influence of individual supernovae (Ryan et al.\ 1996; McWilliam 1997). 
One bonus from this is the case of CS 22892--052 with low metallicity  
and enhanced r-process, enabling a credible thorium chronology to be 
applied (Cowan et al.\ 1999). The incidence of this scatter is 
consistent with the view that stars are formed in globular-cluster 
sized superbubbles of the order of $10^ 5M_{\odot}$, dominated by 
output from a single supernova ($2M_{\odot}$ of oxygen) if the 
oxygen mass fraction in the ISM is under $2\times 10^{-5}$, i.e. 
$2\times 10^{-3}$ of solar. As the metallicity of the ISM increases, 
the influence of an individual supernova is diluted and there is a 
semblance of smooth chemical evolution (Tsujimoto, Shigeyama \& Yoshii 1999). 

\vspace{-2mm} 
\section{Metal supply to the intra-cluster medium}

Hot X-ray gas in clusters of galaxies has a mean metallicity of the order 
of $-0.4$, whether measured in [Fe/H] or [$\alpha$/H], and the mass of 
metals is proportional to that of stars in E and S0 galaxies in the 
cluster (Arnaud et al.\ 1992). As discussed by Renzini et al.\ (1993) and 
Pagel (1997), this requires a large yield of the order of $2Z_{\odot}$ 
if the metals are supplied by stars in the galaxies, reminiscent of 
what comes from the conventional form of the Salpeter IMF, but high 
compared with the yield of $0.7Z_{\odot}$ or so required to fit the 
MDF in the solar neighbourhood (e.g.\ Pagel \& Tautvai\v sien\. e 1995).   
Does this imply a more top-heavy IMF (e.g.\ Arimoto \& Yoshii 1987)? 
Because of the metallicity-luminosity relation (e.g.\ Zaritsky, 
Kennicutt \& Huchra 1994) and considerations of cosmic chemical evolution 
outlined below, I prefer to think of a universal IMF with a high yield, 
modified by outflow from the smaller galaxies.  While an effective 
blowout due to supernova feedback may be difficult to achieve in 
medium-sized galaxies as we see them now (MacLow \& Ferrara 1999), 
there are other mechanisms like tides and ram-pressure stripping, 
and the galaxies that we see today may have been smaller in the past,  
before being built up by inflow or put together by hierarchical 
clustering. 

\vspace{-2mm} 
\section{Cosmic chemical evolution and dark metals} 

Observations at high red-shifts, both of emission from star-forming 
galaxies (Madau et al.\ 1996; Blain et al.\ 1999) and of absorption 
lines in Lyman-$\alpha$ systems (Pettini et al.\ 1999) have led 
to interesting investigations of cosmic chemical evolution (Pei, Fall 
\& Hauser 1999). These may account for only a fraction of the 
metals in the universe, however.  From Big-bang nucleosynthesis, we 
believe that the smoothed-out density of baryonic matter is 
\begin{equation} 
\Omega_{b}h_{70}^ 2 \simeq 0.035 
\end{equation} 
(Tytler et al.\ 2000), a value just consistent within errors 
with recent deductions from BOOMERANG and MAXIMA MWB observations 
(Tegmark \& Zaldarriaga 2000; Balbi et al.\ 2000) or possibly even a slight 
underestimate, whereas the density of stars is only 1/10 as much 
(Fukugita, Hogan \& Peebles 1998). The remainder could be in the form 
of diffuse intergalactic gas, low surface-brightness galaxies, MACHOs 
or something else.  For the first two of these, the metal content is 
certainly an issue.  Mushotzky \& Loewenstein (1997) have argued that 
the intergalactic gas dominates, with the same metallicity as the 
intra-cluster gas, implying a yield of $2.5Z_{\odot}$, while numerical 
simulations by Cen \& Ostriker (1999) imply a somewhat lower metallicity 
like 0.1$Z_{\odot}$ requiring a yield of about $1.5Z_{\odot}$ (cf.\ Pagel 
1999); thus half or more of the heavy elements in the universe are 
as yet unseen, although there is a hint of their presence in recent 
FUSE observations of O {\sc vi} (Tripp et al. 2000). Models of cosmic 
chemical evolution disregard the silent majority of dark metals at their 
peril!

\end{document}